\documentclass[sigconf,anonymous=false]{acmart}

\renewcommand\footnotetextcopyrightpermission[1]{} 
\usepackage{booktabs} 
\DeclareMathSizes{9}{8}{7}{5}
\renewcommand\footnotetextcopyrightpermission[1]{} 
\usepackage[flushleft]{threeparttable}

\usepackage{algorithm}
\usepackage[noend]{algorithmic}
\usepackage{bm}
\usepackage{mathtools}
\usepackage{autonum}
\usepackage{tabularx}
\usepackage{enumitem}
\usepackage{subfig}

\newcolumntype{R}{>{\raggedleft\arraybackslash}X}

\newcommand*{\ta}{\operatornamewithlimits{accuracy}}
\newcommand*{\mqta}{\operatornamewithlimits{\emph{q}-accuracy}}

\newcommand*{\avgprec}{\operatornamewithlimits{avg-prec}}
\newcommand*{\avgrec}{\operatornamewithlimits{avg-rec}}
\newcommand*{\avgef}{\operatornamewithlimits{avg-F1}}

\DeclareMathOperator*{\argmax}{argmax}

\newcommand*{\dir}{\operatorname{Dir}}
\newcommand*{\cat}{\operatorname{Cat}}
\newcommand*{\bern}{\operatorname{Bern}}
\newcommand*{\unif}{\operatorname{Uniform}}
\newcommand*{\pval}{\operatorname{\emph{p}-value}}
\newcommand*{\ttest}{\operatorname{\emph{t}-test}}
\newcommand*{\red}{\textcolor{red}}
\newcommand*{\cntword}{\operatornamewithlimits{T}}
\newcommand*{\cntpc}{\operatornamewithlimits{R}}
\newcommand*{\cntq}{\operatornamewithlimits{U}}

\copyrightyear{2018}
\acmYear{2018}
\setcopyright{acmlicensed}
\acmConference[Woodstock '18]{Woodstock '18: ACM Symposium on Neural Gaze Detection}{June 03--05, 2018}{Woodstock, NY}
\acmBooktitle{Woodstock '18: ACM Symposium on Neural Gaze Detection, June 03--05, 2018, Woodstock, NY}
\acmPrice{15.00}
\acmDOI{10.1145/1122445.1122456}
\acmISBN{978-1-4503-9999-9/18/06}

\begin{document}
\title{Distant-supervised slot-filling for e-commerce queries}
\subtitle{}

\author{Saurav Manchanda}
\affiliation{%
  \institution{University of Minnesota}
  \city{Twin Cities}
  \state{MN, USA}
  \postcode{55455}
}
\email{manch043@umn.edu}

\author{Mohit Sharma}
\authornote{Work was done when Mohit was at WalmartLabs.}
\affiliation{%
  \institution{WalmartLabs}
  \city{Sunnyvale}
  \state{CA}
  \postcode{94086}
}
\email{sharm163@umn.edu}

\author{George Karypis}
\affiliation{%
  \institution{University of Minnesota}
  \city{Twin Cities}
  \state{MN, USA}
  \postcode{55455}
 }
 \email{karypis@umn.edu}

\renewcommand{\shortauthors}{S. Manchanda et al.}

\begin{abstract}
Slot-filling refers to the task of annotating individual terms in a query with the corresponding intended product characteristics (\emph{product type, brand, gender, size, color,} etc.). These characteristics can then be used by a search engine to return results that better match the query's product intent. Traditional methods for slot-filling require the availability of training data with ground truth slot-annotation information. However, generating such labeled data, especially in  e-commerce is expensive and time consuming because the number of slots increases as new products are added. In this paper, we present distant-supervised probabilistic generative models, that require no manual annotation.  The proposed approaches leverage the readily available historical query logs and the purchases that these queries led to, and  also exploit co-occurrence information among the slots in order to identify intended product characteristics. 
We evaluate our approaches by considering both how they affect retrieval performance, as well as how well they classify the slots. 
In terms of retrieval, our approaches achieve better ranking performance (up to $156\%$) over Okapi BM25.
Moreover, our approach that leverages co-occurrence information leads to better performance than the one that does not on both the retrieval and slot classification tasks.

\end{abstract}

%
%



\maketitle

\section{Introduction}

\begin{figure}
\centerline{\fbox{\includegraphics[width=0.9\linewidth]{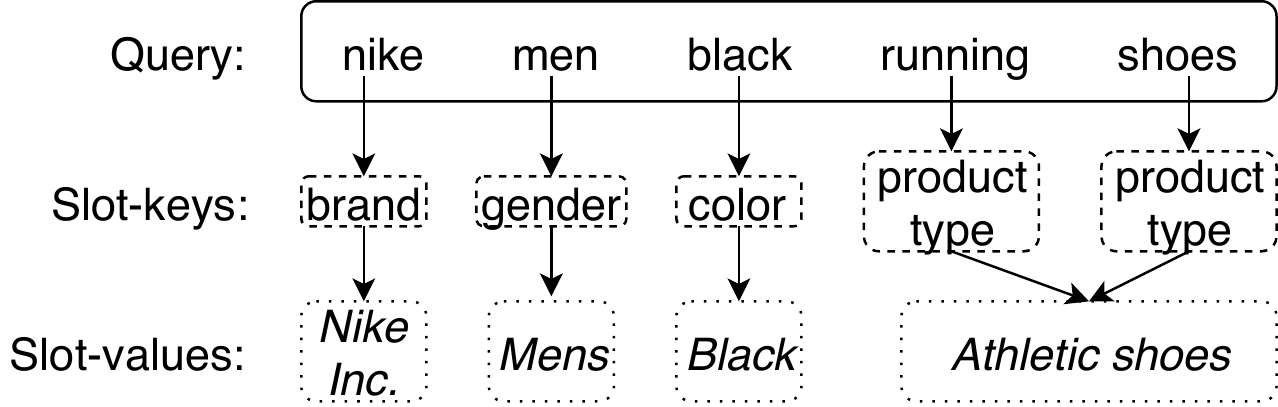}}}
\caption{Query understanding with slot-filling.}
\label{fig:slot_filling_example}
\end{figure}

Online shopping accounts for an ever growing portion of the total retail sales~\footnote{https://www.census.gov/programs-surveys/arts.html}.
%
As such, developing methods to correctly understand the customers' search requirements is an important challenge in e-commerce. Various approaches have been developed to address this challenge, which includes query classification~\cite{cao2009context}, boosting intent-defining terms in the query~\cite{manchanda2019intent, manchanda2019intent2}, to name a few.
Another way to help customers find what they are searching for is to analyze a customer's product search query in order to identify the different product characteristics that the customer is looking for, i.e., \emph{intended product characteristics} such as \emph{product type, brand, gender, size, color, } etc. For example, for the  query ``nike men black running shoes'', the term \emph{nike} describes the \emph{brand}, \emph{men} describes the \emph{gender}, \emph{black} describes the \emph{color} and the terms \emph{running} and \emph{shoes} describe the \emph{product type}. Once the query's intended product characteristics are understood, they can be used by a search engine to return results that correspond to the products whose attributes match these intended characteristics, or as a feature to various Learning to Rank methods~\cite{karmaker2017application}. Slot-filling refers to the task of annotating individual terms in a query with the corresponding intended product characteristics, where each product characteristic is a key-value pair, e.g., \big\{key : \emph{brand}, value : \emph{Nike Inc.}\big\}. Figure \ref{fig:slot_filling_example} illustrates the role of slot-filling in understanding the query's product intent.  


Slot-filling can be thought of as an instance of \emph{entity resolution} or \emph{entity linking}, which is the problem of mapping the textual mentions of entities to their respective entries in a knowledge base. In the case of search queries, these entities are the predefined set of slots (key-value pairs). Slot-filling is traditionally being treated as a word sequence labeling problem, which assigns a tag (slot) to each word in the given input word sequence. 
Traditional approaches require the availability of tagged sequences as the training data~\cite{wang2011semantic, pieraccini1992speech,macherey2001natural,raymond2007generative, wang2005spoken, wang2006discriminative, jeong2008practical, liu2012conversational, jeong2007structures,  xu2013convolutional,mesnil2015using, yao2014spoken, yao2013recurrent, mesnil2013investigation, liu2016attention, vu2016sequential, zhang2016joint}. However, generating such labeled data is expensive and time-consuming because e-commerce is a dynamic domain, where new products continue to be added to the inventory bringing new product-types, brands, etc., into the picture, which leads to a continuously evolving query terms and slot-values. Therefore, labeling datasets or designing heuristics to generate labeled data for e-commerce  is not a one-time thing but has to be done continuously, making the task even more tedious. 
To overcome this problem, approaches have been developed that work in the absence of labeled data~\cite{wang2005spoken, pietra1997fertility, zhou2011learning, henderson2015discriminative}. 
However, these approaches are either specific to the domain of spoken language understanding or assume that the words are very similar to the slot values. These approaches do not work well in cases in which words differ from the slot values, e.g., query ``quaker simply granola'' refers to \emph{cereals} product type without using the term ``cereal''. 

To address the problem of lack of labeled data, we develop \emph{credit attribution} approaches~\cite{ramage2009labeled}, which use engagement data that is readily available in search engine query logs and does not require any manual labeling effort. We present probabilistic generative models that use the products that are engaged (e.g., clicked, added-to-cart, ordered) for the queries in e-commerce search logs as a source of distant-supervision~\cite{mintz2009distant}.
The key insight that we exploit is that in most cases, if a particular query term is associated with products that have a specific value for a slot, e.g., \big\{brand : Nike Inc\big\}, then there is a high likelihood that query term will have that particular slot. Hence, the slots for a query are a subset of the characteristics of the engaged products for that query.
Since our approaches are distant supervision-based, they do not need any information about this subset or mapping of the slots to query words.  Moreover, they also leverage the co-occurrence information of the product characteristics to achieve better performance.

To the best of our knowledge, our work is first of its kind that leverages engagement data in search logs for slot-filling task.
We evaluated our approaches on their impact on the retrieval and on correctly predicting the slots.  
In terms of the retrieval task, our approaches achieve up to $156\%$ better NDCG performance than the Okapi BM25 similarity measure.
Additionally, our approach leveraging the co-occurrence information among the slots gained $\approx 3.3\%$ improvement over the approach that does not leverage the co-occurrence information. With respect to correctly predicting the slots, our approach leveraging the co-occurrence information gained $8\%$ performance improvement over the approach that does not leverage the co-occurrence information. Therefore, the proposed approaches provide an easy solution for slot-filling, by only using the readily available search query logs.

\section{Related work}\label{literature}
The prior research that is most directly related to the work presented in this paper spans the areas of slot filling, entity linking, and credit attribution.

\noindent\emph{\textbf{Slot-filling: }} Slot-filling is a well-researched topic in spoken language understanding, and involves extracting relevant semantic slots from a natural language text. Popular approaches to slot-filling include markov chain methods, conditional random fields and recurrent neural networks.
Generative approaches designed for the slot-filling task includes the ones based on hidden markov models and context free grammar composite models like~\cite{wang2011semantic, pieraccini1992speech, macherey2001natural}. Conditional models designed for slot-filling based on conditional random fields (CRFs) include~\cite{raymond2007generative, wang2005spoken, wang2006discriminative, jeong2008practical, liu2012conversational, jeong2007structures,  xu2013convolutional}. In recent times, recurrent neural networks (RNNs) and convolutional neural networks (CNNs) have been applied to the slot-filling task, and examples of such methods include~\cite{mesnil2015using, yao2014spoken, yao2013recurrent, mesnil2013investigation, liu2016attention, vu2016sequential, zhang2016joint, xu2013convolutional}. A common drawback of these approaches is that they require the availability of tagged sequences as the training data. 
To tackle absence of the labeled training data, some approaches have been developed. These approaches are either specific to the domain of spoken language understanding or make stronger assumptions about the similarity of the words and the slot values~\cite{wang2005spoken, pietra1997fertility, zhou2011learning, henderson2015discriminative}. 
The unsupervised method developed in~\cite{zhai2016query} is closely related to ours but it requires a manually described grammar rules to annotate the queries and is limited to the task of predicting only two slots (product type and brand) as curating grammar for more diverse slots is a non-trivial task.
Furthermore, in e-commerce, new products continue to get added in the inventory bringing new product-types, brands, etc. into picture, which leads to continuously evolving vocabulary. Therefore, designing manual rules to align slot-values to the query words for e-commerce queries requires a continuous manual effort which is expensive.

\noindent\emph{\textbf{Entity linking: }}
Entity linking, also known as record linkage or entity resolution, involves mapping a named-entity to an appropriate entry in a knowledge base.  
Some of the earliest work on entity-linking~\cite{bunescu2006using, cucerzan2005extracting} rely upon the lexical similarity between the context of the named-entity and the features derived from Wikipedia page. The most relevant task to the problems addressed in this paper is the optional entity linking task~\cite{mcnamee2009overview, ji2010overview}), in which the systems can only use the attributes in the knowledge base; this corresponds to the task of updating a knowledge base with no `backing' text, such as Wikipedia text. 
However, our setup is more strict because of the absence of availability of entity attributes and lack of lexical context as most e-commerce queries are concise. 

\noindent\emph{\textbf{Credit attribution: }} 
A document may be associated with multiple labels but all the labels do not apply with equal specificity to the individual parts of the documents. \textit{Credit attribution} problem refers to identifying the specificity of labels to different parts of the document. Various probabilistic and neural-network based approaches have been developed to address the credit-attribution problem, such as Labeled Latent Dirichlet Allocation (LLDA)~\cite{ramage2009labeled}, Partially Labeled Dirichlet Allocation (PLDA)~\cite{ramage2011partially}, Multi-Label Topic Model (MLTM)~\cite{soleimani2017semisupervised}, Segmentation with Refinement (SEG-REFINE)~\cite{manchanda2018text}, and Credit Attribution with Attention (CAWA)~\cite{manchanda2020cawa}. Out of these methods, the ones that are relevant to the methods presented in this paper are the probabilistic graphical approaches (LLDA, PLDA and MLTM), that use topic modelling to associate individual words/sentences in a document with their most appropriate labels.
A common problem with these approaches is that they model the documents as a distribution over the topics and capture the document-level word co-occurrence patterns to reveal topics. These approaches, thus suffer from the severe data sparsity in short documents~\cite{yan2013biterm}. Therefore, these approaches are not suitable for e-commerce queries which tend to be very short. 

\section{Definitions and Notations} \label{definitions}
Let $V$ be the \emph{vocabulary} that corresponds to set of distinct terms that appear across the set of queries $Q$ in a website. The query $q$ is a sequence of $|q|$ terms from the vocabulary $V$, that is,
$q = \langle v_{q_{1}}, \ldots, v_{q_{i}}, \ldots, v_{q_{|q|}}\rangle.$ Each query $q$ is associated with $a_q$ product characteristics (slots), also referred as product intent in e-commerce. The collection of $a_q, \forall q$ is denoted by $A$. Each slot is a key-value pair (e.g., \emph{brand: Nike Inc.}), let $M$ denote the set of all possible slots and $L$ denote the set of all possible slot-keys (e.g., product-type, brand, color etc.).
Let $y_{q_{i}}$ be the slot associated with the term $v_{q_{|i|}}$ in  query $q$ and $y_q$ denotes
the sequence of slots for the terms in the query $q$, i.e., $y_q = \langle y_{q_{1}}, \ldots, y_{q_{|q|}}\rangle$. 
The collection of $y_q, \forall q$ is denoted by $Y$.
Given query $q$ let $c_q$ be the set of slots that is extracted from the characteristics of the products that any user issue $q$ engaged with. The product intent $a_q$ is a subset of $c_q$. 
The set $c_q$ (and hence $a_q$) is constrained to have at most one slot for each unique slot-key, i.e., $c_q$ cannot contain multiple brands, product-types etc. The collection of $c_q, \forall q$ is denoted by $C$.
Let $I$ denote the collection of all possible candidate slots. Throughout the paper, the superscript $-(q,i)$ on a collection symbol denotes that collection excluding the position $i$ in query $q$. For example, $Y^{-(q,i)}$ is the collection of all slot sequences, excluding the one at the  position $i$ of query $q$. Similarly, the superscript $-(q)$ on a collection symbol denotes that collection excluding the query $q$. Star ($\ast$) in place of a symbol indicates a collection with star ($\ast$) taking all possible values. For example, $\phi_*$ denotes the collection of $\phi_i \forall i$. 
Table \ref{tab:notation} provides a reference for the notation used in the paper.

The discrete uniform distribution, denoted by $\unif(\cdot|u)$, has a finite number of values from the set $u$; each value equally likely to be observed with probability $1/|u|$. The Dirichlet distribution, denoted by $\dir(\cdot|\bm{\lambda})$, is parameterized by the vector $\bm{\lambda}$ of positive reals. A $k$-dimensional Dirichlet random variable $\bm{x}$ can take values in the ($k-1$) simplex.
The categorical distribution, denoted by $\cat(\cdot|\bm{x})$, is a discrete probability distribution that describes the possible results of a random variable that can take on one of $k$ possible categories, with the probability of each category separately specified in the vector $\bm{x}$. 
The Dirichlet distribution is the conjugate prior of the categorical distribution. The posterior mean of a categorical distribution with the Dirichlet prior is given by

\begin{equation}\label{eq:cat_posterior}
    \bm{x}_i = \frac{\bm{\lambda}_i + n_i}{\sum_{i=1}^{k}(\bm{\lambda}_i + n_i)}, \forall i\in[1, \ldots,k],
\end{equation}
where, $n_i$ is the number of times category $i$ is observed.
The Bernoulli distribution, denoted by $\bern(\cdot|p)$ is a discrete distribution having two possible outcomes, i.e., \emph{selection}, that occurs with probability $p$, and \emph{rejection}, that occurs with probability $1-p$. 

\begin{table}[!t]
\small
\centering
  \caption{Notation used throughout the paper.}
  \begin{tabularx}{\columnwidth}{lX}
    \hline
Symbol   & Description \\ \hline
$V$    & Vocabulary (set of all the terms). \\
$q$    & A query. \\
$Q$    & Collection of all the queries $q$. \\
$z_q$    & The product category of the query $q$ \\
$Z$    & Collection of $z_q, \forall q$ \\
$a_q$    & The slot-set for the query $q$\\
$A$    & Collection of $a_q, \forall q$ \\
$c_q$    & The candidate slot-set for the query $q$ \\
$C$    & Collection of $c_q, \forall q$ \\
$I$    & Collection of all possible candidate slot-sets \\
$\omega_q$    & A set with $\omega_{q,i} = 1$ if the candidate slot $c_{q,i}$ is present in the slot-set $a_q$, and $0$ otherwise. \\
$\Omega$    & Collection of $\omega_q, \forall q$ \\
$y_q$    & The sequence of slots for the terms in the query $q$ \\
$Y$    & Collection of $y_q, \forall q$ \\
$K$    & Set of all the product categories \\
$M$    & Set of all the possible slots (product characteristics) \\
$L$    & Set of all the possible slot-keys \\
$\bm{\alpha}$    & Dirichlet prior for sampling the product categories.\\
$\bm{\beta}$    & Set of $|K|$ vectors, $\bm{\beta}_k$ denoting the Dirichlet prior for sampling the slots from the product category $k$.\\
$\gamma$    & Bernoulli parameter for selecting/rejecting a slot from $c_q$\\
$\bm{\delta}$    & Set of $|M|$ vectors, $\bm{\delta}_m$ denoting the Dirichlet prior for the query words from the slot $m$.\\
$\bm{\zeta}$    & Set of $|M|$ vectors, $\bm{\zeta}_m$ denoting the Dirichlet prior for transitioning from the slot $m$.\\
$\bm{\phi}$    & Probability distribution of the product categories in the query collection\\
$\bm{\chi}$    & Set of $|K|$ vectors, $\bm{\chi}_k$ denoting the probability distribution of the slots in the product category $k$.\\
$\bm{\psi}$    & Set of $|M|$ vectors, $\bm{\psi}_m$ denoting the probability distribution of the query words in the slot $m$.\\
$\bm{\upsilon}$    & Set of $|M|$ vectors, $\bm{\upsilon}_m$ denoting the transition probability distribution from the slot $m$.\\
\hline
\end{tabularx}
  \label{tab:notation}
\end{table}

\section{Proposed methods}\label{proposed}

In order to address the problem of finding the most appropriate slot for the terms in a query, when the labeled training data is unavailable, we developed generative probabilistic approaches. Our approaches assume that the product intent of a query correspond to a subset of the product characteristics of the engaged products, and leverage the information of the engaged products from the historical search logs as a source of distant-supervision. The training data for our approaches are the search queries and the product characteristics of the engaged products that form the corresponding candidate slot-sets. For example, for the search query ``nike men running shoes'', the characteristics of an engaged product are \{\emph{product-type: athletic shoes, brand: Nike Inc., color: red, size: 10, gender: mens}\}. These characteristics form the candidate slot-set for the search query.
Given a query $q$ and its candidate slot-set $c_q$, our approaches find the most probable slot-sequence $y_q$, such that $y_{q,i} \in c_q, \forall i$. Our approaches achieve this by maximizing the joint probability of the observed query words, observed candidate slots and the unobserved slot-sequence.

 \begin{algorithm}[!t]
 \caption{Outline of the generative process for slot-filling}
 \begin{algorithmic}[1]
  \FOR {each query $q \in Q$}
    \STATE Generate $c_q$  from I;  $a_q$  from $c_q$; $y_{q}$ from $a_q$
    \FOR {each $i = 1$ to $|q|$}
        \STATE Generate $v_{q,i}$ from ${y_{q,i}}$
        
    \ENDFOR
        
  \ENDFOR
 \end{algorithmic}
 \label{alg:generative_all}
 \end{algorithm}
Our approaches follow some variation of the common generative process outlined in Algorithm~\ref{alg:generative_all}. The generative process has four unknowns: generating the candidate slot-set $c_q$ from $I$, generating the product intent $a_q$ from $c_q$, generating the slot sequence $y_{q}$ from $a_q$ and generating each query word $v_{q,i}$ from $y_{q,i}$. To generate $v_{q,i}$ from $y_{q,i}$, our approaches model each slot $m$ as a categorical distribution over the vocabulary $V$. The categorical distribution for the slot $m$, denoted by $\bm{\psi}_m$, is sampled from a Dirichlet distribution with prior $\bm{\delta}_{m}$. Our approaches differ in the way they model the other generative steps.






Our first approach, \emph{uniform slot distribution (USD)}, arguably the simplest approach, assumes that $c_q$ is sampled from the $I$ under a discrete uniform distribution. USD directly generates $y_q$ from $c_q$, thus bypassing generating $a_q$. USD samples each slot $y_{q,i}$ in the slot-sequence independently from a uniform discrete distribution.

The rest of our approaches relax the independence assumptions of the USD and leverage different ideas to capture the co-occurrence of the slots. \emph{Markovian slot distribution (MSD)} leverages the idea that two co-occurring slots ($m_1$ and $m_2$) should have high transition probabilities, i.e., probability of $m_2$ followed by $m_1$ and/or $m_1$ followed by $m_2$ should be relatively high. \emph{Correlated uniform slot distribution (CUSD)} extends USD so as to sample $c_q$ from $I$ such that the slots in $c_q$ are more probable to co-occur. To model this co-occurrence, CUSD leverages the idea that the product intent of a search query is sampled from one of the product categories (set of semantically similar product characteristics). 
As opposed to CUSD which models all the slots in $c_q$ as belonging to the same product category, \emph{correlated uniform slot distribution with subset selection (CUSDSS)} assumes that a subset of $c_q$ belong to a product category, and this subset corresponds to the actual product intent of the query ($a_q$). For example, consider a laundry detergent product with the following slots: \{\emph{product-type: laundry detergent, brand: tide, color: yellow}\}, but we expect that \emph{color} is not an important attribute when someone is buying laundry detergent. 

In the remainder of this section, we discuss these approaches. Because of space constraints, we only provide final learning and inference equations for each of the approaches, and provide detailed generative process and associated derivation in the attached appendix.


\begin{figure}%
\centering
\subfloat[USD]{\includegraphics[width=0.5\linewidth]{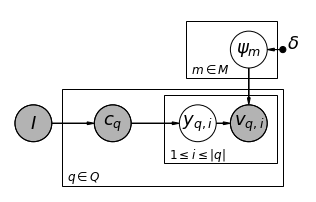}}
\subfloat[MSD]{\includegraphics[width=0.5\linewidth]{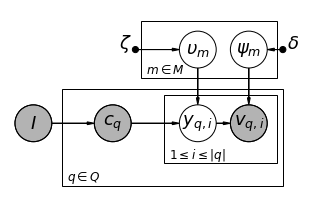}}\\
\subfloat[CUSD]{\includegraphics[width=0.5\linewidth]{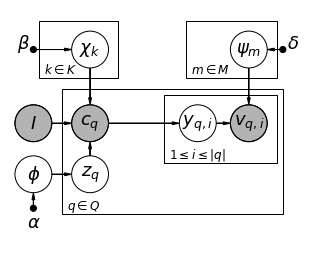}}
\subfloat[CUSDSS]{\includegraphics[width=0.5\linewidth]{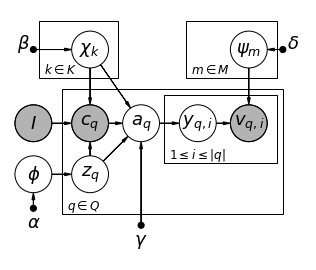}}
\caption{Plate notation of the proposed approaches}
\label{fig:plate}
\end{figure} 
 
\subsection{Uniform Slot Distribution (USD)}

 \begin{algorithm}[!t]
 \caption{Generative process for USD}
 \begin{algorithmic}[1]
  
  \FOR {each characteristic $m \in M$}
    \STATE Generate $\bm{\psi}_m = \langle \psi_{m,1}, \ldots, \psi_{m,|V|}\rangle \sim \dir(\cdot|\bm{\delta_{m}})$
  \ENDFOR
  \FOR {each query $q \in Q$}
        \STATE Generate $c_q \sim \unif(\cdot|I)$
  
    \FOR {each $i = 1$ to $|q|$}
        \STATE Generate $y_{q,i} \sim \unif(\cdot|c_q)$
        \STATE Generate $v_{q,i} \sim \cat(\cdot|\bm{\psi}_{y_{q,i}})$
        
    \ENDFOR
        
  \ENDFOR
 \end{algorithmic}
 \label{alg:generative_usd}
 \end{algorithm}
The \emph{uniform slot distribution (USD)} model assumes that $c_q$ is sampled from the $I$ under a uniform distribution. Further, each slot is sampled independently from a uniform distribution i.e.,
\begin{equation}
    P(y_q|c_q) = \prod_{i=1}^{|q|}P(y_{q,i}|c_q) = \prod_{i=1}^{|q|}\frac{1}{|c_q|}.
\end{equation}
   
Algorithm~\ref{alg:generative_usd} and Figure~\ref{fig:plate}(a) shows the generative process for plate notation for the USD, respectively.

\subsubsection{Learning: }As mentioned before, our approaches model each slot as a categorical distribution over the vocabulary $V$. The probability distribution for the slot $m$ is denoted by $\bm{\psi}_m$, and is sampled from a Dirichlet distribution with prior $\bm{\delta}_{m}$.
The overall joint probability distribution, according the USD model is given by:
\begin{equation}
    P(Q,Y,C,\bm{\psi}_{*}|\bm{\delta}_{*}) = \prod_{m \in M}P(\bm{\psi}_m|\bm{\delta}_{m})
    \prod_{q \in Q}\frac{1}{|I|}\prod_{i=1}^{|q|}\frac{1}{|c_q|}P(v_{q,i}|\bm{\psi}_{y_{q,i}})
\end{equation}
\begin{equation}
     = \prod_{m \in M}\dir(\bm{\psi}_m|\bm{\delta}_{m})
    \prod_{q \in Q}\frac{1}{|I|}\prod_{i=1}^{|q|}\frac{1}{|c_q|}\cat(v_{q,i}|\bm{\psi}_{y_{q,i}})
\end{equation}
We use Gibbs sampling to perform approximate inference. 
We estimate the parameters iteratively, by sampling a parameter from its distribution conditioned on the values of the remaining parameters. 
The parameters that we need to estimate are $\bm{\psi}$ and $Y$. However, since Dirichlet is the conjugate prior of the categorical distribution, we can integrate out $\bm{\psi}$ and use collapsed Gibbs sampling~\cite{griffiths2004finding} to only estimate the $Y$. Once the inference is complete, $\bm{\psi}$ can then be computed from $Y$. Therefore, we only have to sample the slot for each query word from its conditional distribution given the remaining parameters, the probability distribution for which is given by

\begin{equation}\label{eq:uds_posterior}
P(y_{q,i}|Q, C, Y^{-(q,i)},\bm{\delta}_{*}) \propto \frac{\bm{\delta}_{y_{q,i}, v_{q,i}}  + \cntword^{-(q,i)}(y_{q,i}, v_{q,i})}{\sum_{v' \in V}(\bm{\delta}_{y_{q,i}, v'}  + \cntword^{-(q,i)}(y_{q,i}, v')},
\end{equation}
where, $\cntword(a, b)$  is the count of word $a$ tagged with the slot $b$. The RHS in Equation~\eqref{eq:uds_posterior} is exactly the posterior distribution of $\bm{\psi}_{y_q}$, excluding the current assignment. Once the learning is complete, $\bm{\psi}_{m}, \forall m$ can be estimated as in Equation~\eqref{eq:cat_posterior}.

\subsubsection{Inference: }The slot for a query word $v_{q,i}$ is simply the slot $m$ for which $\bm{\psi}_{m,v_{q,i}}$ is maximum, i.e.,$
    y_{q,i} = \argmax_{m\in c_q} \bm{\psi}_{m,v_{q,i}}$.

\subsection{Markovian Slot Distribution (MSD)}

 \begin{algorithm}[!t]
 \caption{Generative process for MSD}
 \begin{algorithmic}[1]
  
  \FOR {each characteristic $m \in M$}
    \STATE Generate $\bm{\psi}_m = \langle \psi_{m,1}, \ldots, \psi_{m,|V|}\rangle \sim \dir(\cdot|\bm{\delta}_{m})$
    \STATE Generate $\bm{\upsilon}_m = \langle \upsilon_{m,1}, \ldots, \upsilon_{m,|M|}\rangle \sim \dir(\cdot|\bm{\zeta}_{m})$
  \ENDFOR
  \FOR {each query $q \in Q$}
    \STATE Generate $c_q \sim \unif(\cdot|I)$
  
    \FOR {each $i = 1$ to $|q|$}
        \STATE Generate $y_{q,i} \sim\cat(\cdot|\bm{\upsilon}_{y_{q,i-1}})$; $v_{q,i} \sim \cat(\cdot|\bm{\psi}_{y_{q,i}})$
        
    \ENDFOR
        
  \ENDFOR
 \end{algorithmic}
 \label{alg:generative_msd}
 \end{algorithm}
 
Similar to the USD, the \emph{markovian slot distribution (MSD)} model samples $c_q$ from $I$ under a discrete uniform distribution. But, instead of sampling each slot independent of the other slots, MSD assumes that the slots in the slot sequence $y_q$ follow a first order markov process, i.e.,
\begin{equation}
  P(y_q|c_q) = \prod_{i=1}^{|q|}P(y_{q,i}|c_q) = \prod_{i=1}^{|q|}P(y_{q,i}|y_{q,i-1}).
\end{equation}

MSD leverages the idea that two co-occurring slots ($m_1$ and $m_2$) should have high transition probabilities, i.e., $P(m_2|m_1)$ and/or $P(m_2|m_1)$ should be high.
 The generative process for MSD is outlined in Algorithm~\ref{alg:generative_msd} and the plate notation is shown in Figure~\ref{fig:plate}(b).
\subsubsection{Learning: }To learn the transition probability $P(m_1|m_2)$, we model each slot as a categorical distribution over the set of all slots $M$. The probability distribution for the slot $m$ is denoted by $\bm{\upsilon}_m$, and is sampled from a Dirichlet distribution with prior $\bm{\zeta}_{m}$.
The joint probability distribution, according to the MSD model is given by:
\begin{multline}
    P(Q,Y,C,\bm{\psi},\bm{\upsilon}_{*}|\bm{\delta}_{*},\bm{\zeta}_{*}) = \prod_{m \in M}P(\bm{\psi}_m|\bm{\delta}_{m})\prod_{m \in M}P(\bm{\upsilon}_m|\bm{\zeta}_{m})\\
    \prod_{q \in Q}\frac{1}{|I|}\prod_{i=1}^{|q|}P(y_{q,i}|\bm{\upsilon}_{y_{q,i-1}})P(v_{q,i}|\bm{\psi}_{y_{q,i}})
\end{multline}

\begin{multline}
    = \prod_{m \in M}\dir(\bm{\psi}_m|\bm{\delta}_{m})\prod_{m \in M}\dir(\bm{\upsilon}_m|\bm{\zeta}_{m})\\
    \prod_{q \in Q}\frac{1}{|I|}\prod_{i=1}^{|q|}\frac{\cat(y_{q,i}|\bm{\upsilon}_{y_{q,i-1}})}{\sum_{m\in c_q}\cat(m|\bm{\upsilon}_{y_{q,i-1}})}\cat(v_{q,i}|\bm{\psi}_{y_{q,i}})
\end{multline}

Note the important distinction between the MSD model and the usual hidden markov model, that we are only sampling the states (slots) from the $c_q$ and not $M$. Therefore, $P(y_{q,i}|\bm{\upsilon}_{y_{q,i-1}}) != \cat(y_{q,i}|\bm{\upsilon}_{y_{q,i-1}})$. This makes integrating out $\bm{\upsilon}_{*}$ difficult. Hence, we don't collapse out $\bm{\upsilon}_{*}$, but estimate it in course of our learning. The conditional distribution for the slot of a query word, given the remaining parameters, is given by:

\begin{equation}
P(y_{q,i}|Q, C, Y^{-v_{q,i}},\bm{\upsilon}_{*},\bm{\delta}_{*},\bm{\zeta}_{*}) \propto P1 \times P2,
\end{equation}
where $P1$ corresponds to sampling the slot $y_{q,i}$ and equals,
\begin{equation}
P1 = \bm{\upsilon}_{y_{q,i-1}, y_{q,i}} \times \bm{\upsilon}_{y_{q,i}, y_{q,i+1}}.
\end{equation}
$P2$ corresponds to sampling the query word $v_{q,i}$ from the slot $y_{q,i}$ and equals
\begin{equation}
P2 = \frac{\bm{\delta}_{y_{q,i}, v_{q,i}}  + \cntword^{-(q,i)}(y_{q,i}, v_{q,i})}{\sum_{v' \in V}(\bm{\delta}_{y_{q,i}, v'}  + \cntword^{-(q,i)}(y_{q,i}, v')}.
\end{equation}
Equation~\eqref{eq:cat_posterior} is used to estimate $\bm{\psi}_{*}$ and $\bm{\upsilon}_{*}$.

\subsubsection{Inference: }Once we have $\bm{\psi}_{*}$ and $\bm{\upsilon}_{*}$, the slot-sequence for a query can be estimated using the Viterbi decoding algorithm~\cite{viterbi1967error}.

The primary advantage of exploiting the co-occurrence is to perform selection of $c_q$ from $I$, when $c_q$ is not observed. For this, we can use the transition probabilities of the slot sequence as a proxy of how much the slots are likely to co-occur. To achieve this, we estimate the slot sequence, and the corresponding candidate slot set as below:

\begin{equation}
    y_q, c_q = \argmax_{\hat{y}_q, \hat{c}_q} (P(\hat{y}_q|\hat{c}_q))^\mu P(\hat{y}_q|q,\hat{c}_q); \hat{c}_q \in I,
\end{equation}
where, $P(\hat{y}_q|\hat{c}_q)) = \prod_{i=1}^{|q|-1}P(y_{q,i+1}|y_{q,i})$, $\mu$ is another hyperparameter establishing how much co-occurrence information we want to incorporate and $P(\hat{y}_q|q,\hat{c}_q)$ corresponds to a posteriori estimate.

\subsection{Correlated Uniform Slot Distribution (CUSD)}
 \begin{algorithm}[!t]
 \caption{Generative process for sampling a product category and intended slots from the product category}
 \begin{algorithmic}[1]
    \STATE Generate $\bm{\phi} = \langle \phi_{1}, \ldots, \phi_{|K|}\rangle \sim \dir(\cdot|\bm{\alpha})$
  \FOR {each product category $k \in K$}
    \STATE Generate $\bm{\chi}_k = \langle \chi_{k,1}, \ldots, \chi_{k,|M|}\rangle \sim \dir(\cdot|\bm{\beta}_{k})$
  \ENDFOR
  \FOR {each query $q \in Q$}
    \STATE Generate $z_q  \sim \cat(\cdot|\bm{\phi})$
  
    \FOR {each $i = 1$ to $|c_q|$}
        \STATE Generate $c_{q,i} \sim \cat(\cdot|\bm{\chi}_{z_{q}})$
        
    \ENDFOR
        
  \ENDFOR
 \end{algorithmic}
 \label{alg:generative_category}
 \end{algorithm}
 To model the co-occurrence among the slots, CUSD leverages the idea that slots in the candidate slot-set $c_q$ is sampled from one of the product categories.  CUSD extends USD by favoring that $c_q$ such that the slots in $c_q$ are more probable of belonging to the same product category. The number of product categories is a hyperparameter denoted by $K$. Hence, sampling $c_q$ from $I$ corresponds to sampling a product category and then sampling the slot set $c_q$ from this product category. This sampling step is modeled as a mixture of unigrams, i.e., $P(c_q|I) = P(z_q)\prod_{i=1}^{|q|}P(c_{q,i}|z_q)$.

 The generative process for sampling the product category and the candidate slots from the product category is outlined in Algorithm~\ref{alg:generative_category}. 
 The plate notation for CUSD is shown in Figure~\ref{fig:plate}(c). 
Note that, when $c_q$ is observed (training queries), the slot annotations generated by USD and CUSD are the same. 

\subsubsection{Learning: }We denote the probabilty of sampling a product category ($P(z_q)$) as $\bm{\phi}$, and model it as a categorical distribution sampled from a Dirichlet distribution with prior $\bm{\alpha}$. Each product category is further modeled as categorical distribution over the slots $M$. The probability distribution for the product category $k$ over the slots is denoted by $\bm{\chi}_k$, and is sampled from a Dirichlet distribution with prior $\bm{\beta}_{k}$. For the training data, since the candidate slot sets are observed, the candidate slot set sampling step can be treated independently of the further sampling steps (same steps as the USD). So, we only look into how to estimate the parameters corresponding to the product categories. The overall joint probability distribution of sampling the candidate slot set is:
\begin{equation}
    P(Z,C,\bm{\phi}_{*},\bm{\chi}_{*}|\bm{\alpha},\bm{\beta}_{*}) = P(\bm{\phi}|\bm{\alpha})\prod_{k \in K}P(\bm{\chi}_k|\bm{\beta}_{k})
    \prod_{q \in Q}P(z_q|\bm{\phi})\prod_{i=1}^{|c_q|}P(c_{q,i}|\bm{\chi}_{z_q})
\end{equation}

\begin{equation}
    = \dir(\bm{\phi}|\bm{\alpha})\prod_{k \in K}\dir(\bm{\chi}_k|\bm{\beta}_{k})
    \prod_{q \in Q}\cat(z_q|\bm{\phi})\prod_{i=1}^{|c_q|}\cat(c_{q,i}|\bm{\chi}_{z_q})
\end{equation}

We can integrate out the $\bm{\phi}_{*}$ and $\bm{\chi}_{*}$, giving the update equations for the product category of a query as:

\begin{equation}
P(z_{q}|C, Z^{-(q)},\bm{\alpha},\bm{\beta}_{*}) \propto P1 \times P2,
\end{equation}
where $P1$ corresponds to sampling a product category and equals
\begin{equation}
P1 = \bm{\alpha}_{z_q}  + \cntq^{-(q)}(z_q),
\end{equation}
where, $\cntq(z_q)$ is the count of the queries sampled from the product category $z_q$. 
$P2$ corresponds to sampling each of the slot in $c_q$ from the product category $z_q$ and equals
\begin{equation}
P2 = \frac{\prod_{m \in c_q} (\bm{\beta}_{z_q, m}  + \cntpc^{-(q)}(z_q, m))}{ \prod_{i=0}^{|c_q|-1}(\sum_{m' \in M}(\bm{\beta}_{z_q, m'}  + \cntpc^{-(q)}(z_q, m')) + i)},
\end{equation}
where $\cntpc(z_q, m)$ is the count of the slot $m$ sampled from $z_q$. 

\subsubsection{Inference: }The product category for a query $q$ given its candidate slot set $c_q$ can be found using a maximum a posteriori estimate, i.e., $
    z_{q} = \argmax_{k} \bm{\phi}_{k}\prod_{i=1}^{|c_q|}\bm{\chi}_{k,c_{q,i}}$. The advantage of CUSD over the USD is to perform selection of $c_q$ from $I$, when $c_q$ is not observed. For this, we can use the probability of the candidate slot-set $c_q$ of being sampled from the product category $z_q$ as a proxy of how much the candidate slots are likely to co-occur. To achieve this, we estimate the slot sequence, and the corresponding candidate slot set as below:

\begin{equation}\label{eq:c_est}
    y_q, c_q = \argmax_{\hat{y}_q, \hat{c}_q} (P(\hat{c}_q, \hat{z}_q))^\mu P(\hat{y}_q|q,\hat{c}_q); \hat{c}_q \in I,
\end{equation}
where, $P(\hat{c}_q, \hat{z}_q) = \bm{\phi}_{\hat{z}_q}\prod_{i=1}^{|\hat{c}_q|}\bm{\chi}_{\hat{z}_q,\hat{c}_{q,i}}$, $\mu$ is another hyperparameter establishing how much co-occurrence information we want to incorporate and $P(\hat{y}_q|q,\hat{c}_q)$ corresponds to a posteriori estimate according to USD. 

\subsection{Correlated Uniform Slot Distribution with Subset Selection (CUSDSS)}

 \begin{algorithm}[!t]
 \caption{Generative process for CUSDSS}
 \begin{algorithmic}[1]
    \STATE Generate $\bm{\phi} = \langle \phi_{1}, \ldots, \phi_{|K|}\rangle \sim \dir(\cdot|\bm{\alpha})$
  \FOR {each product category $k \in K$}
    \STATE Generate $\bm{\chi}_k = \langle \chi_{k,1}, \ldots, \chi_{k,\Bar{M}}\rangle \sim \dir(\cdot|\bm{\beta}_{k})$
  \ENDFOR
  
  \FOR {each slot $m \in M$}
    \STATE Generate $\bm{\psi}_m = \langle \psi_{m,1}, \ldots, \psi_{m,|V|}\rangle \sim \dir(\cdot|\bm{\delta}_{m})$
  \ENDFOR
  \FOR {each query $q \in Q$}
    \STATE Generate $z_q  \sim \cat(\cdot|\bm{\phi})$
    \FOR {each slot $i = 1$ to $|c_q|$}
        \STATE Generate $\omega_{q,i} \in \{0,1\} \sim \bern(\cdot|\gamma)$
    \ENDFOR
    \FOR {each $i = 1$ to $|c_q|$}
    
        \IF{$\omega_{q,i} == 1$}
            \STATE Generate $c_{q,i} \sim \cat(\cdot|\bm{\chi}_{z_{q}})$; Add $c_{q,i}$ to $a_q$
        \ELSE
            \STATE Generate $\neg c_{q,i} \sim \cat(\cdot|\bm{\chi}_{z_{q}})$
        \ENDIF
        
    \ENDFOR
    \FOR {each $i = 1$ to $|q|$}
        \STATE Generate $y_{q,i} \sim \unif(\cdot|a_q)$; $v_{q,i} \sim \cat(\cdot|\bm{\psi}_{y_{q,i}})$
        
    \ENDFOR
        
  \ENDFOR
 \end{algorithmic}
 \label{alg:generative_subset}
 \end{algorithm}

CUSD leverages the co-occurrence among the slots by favoring $c_q$ such that the slots in $c_q$ are more probable of belonging to the same product category. However, not all the slots in $c_q$ are related to each other. For example, the query \emph{tide detergent liquid} has the following product intent : \{\emph{product-type: laundry detergent, brand: tide}\}, but the engaged product with this query can have more slots than the query's intent, for example \{\emph{color: yellow}\}, as the catalog records all the possible slots. We expect that the \emph{color} is not an important attribute when someone is buying detergent. In this case, we expect that the slot \emph{color: yellow} is not related to the other slots in $c_q$. Therefore, a better way to capture the co-occurrence among the slots is the favor that $a_q$ (product intent of the query $q$) such that the slots in $a_q$ are more probable of belonging to the same product category. We present \emph{correlated uniform slot distribution with subset selection (CUSDSS)}, which leverages this idea. CUSDSS extends CUSD by selecting or rejecting each slot in the set $c_q$ using a Bernoulli coin toss for each slot with parameter $\gamma$. The selected slots make up the set $a_q \subseteq c_q$. For each query $q$, we define the set $\omega_{q}$ of size $c_q$, with $\omega_{q,i} = 1$ if $c_{q,i}\in a_q$ and $0$ otherwise. The collection of all these $\omega_{q}, \forall q$ is denoted by $\Omega$. 

The generative process for CUSDSS is outlined in Algorithm~\ref{alg:generative_subset} and the plate notation is shown in Figure~\ref{fig:plate}(d).

\subsubsection{Learning: }Unlike CUSD, we are not sampling the slot-sequence $y_q$ from the observed $c_q$ but from the unobserved $a_q$. Therefore, sampling $a_q$ is not independent of the subsequent sampling steps. The joint probability distribution according to the CUSDSS is:
\begin{multline}
    P(Q,Y,Z,C,A,\bm{\chi}_{*},\bm{\psi}_{*},\bm{\phi},\Omega,\bm{\upsilon}_{*}|\bm{\alpha},\bm{\beta}_{*},\gamma,\bm{\delta}_{*},\bm{\zeta}_{*}) \\= P(\bm{\phi}|\bm{\alpha})\prod_{k \in K}P(\bm{\chi}_k|\bm{\beta}_{k}) \prod_{m \in M}P(\bm{\psi}_m|\bm{\delta}_{m})\prod_{m \in M}P(\bm{\upsilon}_m|\bm{\zeta}_{m})\\
    \prod_{q \in Q}
    \Bigg( P(\omega_q|\gamma) P(z_q|\bm{\phi}) 
    \prod_{i =  1}^{|c_q|} P(c_{q,i}|\bm{\chi}_{z_{q}})^{\omega_{q,i}}P(\neg c_{q,i}|\bm{\chi}_{z_{q}})^{(1 - \omega_{q,i})}\\
    \prod_{i =  1}^{|q|}P(y_{q,i}|\bm{\upsilon}_{y_{q,i-1}}) P(v_{q,i}|\bm{\psi}_{y_{q,i}}) \Bigg)
\end{multline}

An important unknown in the above distribution is how to model $P(\neg c_{q,i}|\bm{\chi}_{z_{q}})$. We model this by introducing a dummy slot $\neg x$ for every slot $x$. We call this extended slot set as $\Bar{M}$. Hence, each product category $k$ is now a distribution over $|\Bar{M}| = 2\times|M|$ slots. 
\begin{multline}
    P(Q,Y,Z,C,A,\bm{\chi}_{*},\bm{\psi}_{*},\bm{\phi},\Omega,\bm{\upsilon}_{*}|\bm{\alpha},\bm{\beta}_{*},\gamma,\bm{\delta_{*}},\bm{\zeta}_{*}) \\= \dir(\bm{\phi}|\bm{\alpha})\prod_{k \in K}\dir(\bm{\chi}_k|\bm{\beta}_{k}) \prod_{m \in M}\dir(\bm{\psi}_m|\bm{\delta}_{m})\prod_{m \in M}\dir(\bm{\upsilon}_m|\bm{\zeta}_{m})\\
    \prod_{q \in Q}
    \Bigg( \cat(z_q|\bm{\phi}) \prod_{i = 1}^{|c_q|}\gamma^{\omega_{q,i}}(1 - \gamma)^{(1 - {\omega_{q,i}})}\\
    \prod_{i =  1}^{|c_q|} \cat(c_{q,i}|\bm{\chi}_{z_{q}})^{\omega_{q,i}}\cat(\neg c_{q,i}|\bm{\chi}_{z_{q}})^{(1 - \omega_{q,i})}\\
    \prod_{i =  1}^{|q|}\frac{\cat(y_{q,i}|\bm{\upsilon}_{y_{q,i-1}})}{\sum_{m\in a_q}\cat(m|\bm{\upsilon}_{y_{q,i-1}})} \cat(v_{q,i}|\bm{\psi}_{y_{q,i}}) \Bigg)
\end{multline}

We can integrate out the $\bm{\chi}_{*}, \bm{\psi}_{*}$ and $\bm{\phi}_{*}$. We then perform block Gibbs sampling updating $a_q$, $z_q$ and $y_{q,i}$ in one step, with the update equation as:
\begin{multline}
P(y_{q,i},z_{q},a_{q}, \omega_q|Q, C, A^{-(q,i)}, Z^{-(q)},\Omega^{-(q,i)}, \bm{\alpha},\bm{\beta}_{*},\gamma,\bm{\delta}_{*}, \bm{\zeta}_{*}) \propto\\
P1 \times P2 \times P3 \times P4 \times P5,
\end{multline}
where, $P1$ corresponds to sampling the product category $z_q$ and equals,
\begin{equation}
P1 = \bm{\alpha}_{z_q}  + \cntq^{-(q)}(z_q).
\end{equation}
$P2$ corresponds to sampling slots in $a_q$ from the product category $z_q$ and equals,
\begin{equation}
P2 =
\frac{\prod_{m \in a_q} (\bm{\beta}_{z_q, m}  + \cntpc^{-(q)}(z_q, m))\prod_{m \in c_q - a_q} (\bm{\beta}_{z_q, m}  + \cntpc^{-(q)}(\neg z_q, m))}{ \prod_{i=0}^{|c_q|-1}(\sum_{m' \in \Bar{M}}(\bm{\beta}_{z_q, m'}  + \cntpc^{-(q)}(z_q, m')) + i)}
\end{equation}
$P3$ corresponds to the subset selection and equals,
\begin{equation}
P3 = \begin{cases}
1 &y_{q,i}\in a_{q-v_{q,i}}\\
\frac{\gamma}{1 -\gamma} &y_{q,i}\notin a_{q-v_{q,i}}
\end{cases}
\end{equation}
$P4$ corresponds to sampling each slot and equals,
\begin{equation}\label{eq:label_bias}
P4 =\frac{1}{|a_q|}.
\end{equation}
$P5$ corresponds to the probability of sampling the query word $v_{q,i}$ given the slot $y_{q,i}$ and equals,
\begin{equation}
P5 = \frac{\bm{\delta}_{y_{q,i}, v_{q,i}}  + \cntword^{-(q,i)}(y_{q,i}, v_{q,i})}{\sum_{v' \in V}(\bm{\delta}_{y_{q,i}, v'}  + \cntword^{-(q,i)}(y_{q,i}, v')}.
\end{equation}

Note that, for subset selection, $\gamma$ doesn't have to necessarily be less than $0.5$. This is because of the \emph{label-bias problem}~\cite{lafferty2001conditional}, which leads to preferring small sized $a_q$. This can also be seen through Equation~\eqref{eq:label_bias}, which causes $P4$ to have higher values for smaller $|a_q|$. Therefore, $\gamma$ can be more than $0.5$ to select the required subset.

\subsubsection{Inference: }We perform the inference iteratively using block Gibbs sampling in the same manner as learning.
Similar to CUSD, we use the probability of the product intent $a_q$ of being sampled from $z_q$ as a proxy of how much the slots in the product intent are likely to co-occur. Therefore, in this case, we estimate the slot sequence, and the corresponding candidate slot set as below:
\begin{equation}\label{eq:ss_est}
    y_q, c_q = \argmax_{\hat{y}_q, \hat{c}_q} (P(\hat{c}_q, \hat{z}_q))^\mu P(\hat{y}_q|q,\hat{c}_q); \hat{c}_q \in I,
\end{equation}
where $P(\hat{c}_q, \hat{z}_q) = \bm{\phi}_{\hat{z}_q}\prod_{i=1}^{|\hat{c}_q|}\bm{\chi}_{\hat{z}_q,\hat{c}_{q,i}}^{\omega_{q,i}}(1 - \bm{\chi}_{\hat{z}_q,\neg\hat{c}_{q,i}})^{(1 - \omega_{q,i})}$ and $\mu$ controls how much co-occurrence information we want to incorporate. 

\section{Experimental methodology}\label{experiments}

\subsection{User engagement and annotated dataset}\label{dataset}

We used two datasets to estimate and evaluate our models. We call the first dataset \emph{engagement}, and it contains pairs of the form $(q, c_q)$, i.e., queries along with their candidate slot-sets. We call the second dataset \emph{annotated} and it contains ground-truth annotated queries. We first describe our process of generating the \emph{engagement} dataset and then describe the \emph{annotated} dataset. 

We used the historical search logs of an e-commerce retailer to prepare our \emph{engagement} dataset. This data contains $48{,}785$ distinct queries, and the size of vocabulary is $1{,}936$. We assume that $c_q$ corresponds to the product characteristics of the products that were engaged for $q$. The product characteristics belong to the following six types (slot-keys) and their number of distinct values is in parenthesis: \emph{product-type} ($983$), \emph{brand} ($807$), \emph{gender} ($7$), \emph{color} ($20$), \emph{age} ($86$) and \emph{size} ($243$).  
We created the required dataset by sampling query-item pairs from the historical logs spanning 13 months (July'17 to July'18), such that the number of orders of that item for that query over a month is at least five. 
Note that each query can be engaged with multiple items, all of which were included to create $c_q$.
We limited our dataset to the queries whose words are frequent, i.e., each query word occurs at least 50 times. Furthermore, we use the items only if their product characteristics are present in at least 50 items. These filtering steps avoid data sparsity, thus, help estimating models accurately.

The \emph{annotated} dataset contains a manually annotated representative sample of the queries.  
The instructions for slot-key annotation along with a few examples illustrating the same were given to annotators. Two annotators tagged each query, and when the annotators disagree over their annotations, they were asked to resolve conflicts through reasoning and discussions. Furthermore, we discarded the queries for which annotators failed to resolve conflicts. 
In addition to the slot-keys \emph{product-type}, \emph{brand}, \emph{gender}, \emph{color}, \emph{age} and \emph{size}, the set of slot-keys contain an additional tag \emph{miscellaneous} corresponding to the words which do not fit into either of the other tags. The final \emph{annotated} dataset contains a total of $3{,}430$ annotated queries.

For reproducibility, the ideal way is to perform training and evaluation on publicly available datasets. However, to the best of our knowledge, there exists no public dataset containing e-commerce queries and their engagement with products having attributes (slot-sets). Hence, we limit our evaluation to the above mentioned proprietary dataset. 

\subsection{Evaluation methodology}
We evaluate our methods on three tasks, as described below:

\noindent\emph{\textbf{T1: }} This task evaluates how well the predicted slots for a query improve the retrieval performance. We calculate the similarity between a query and a product by simply counting the number of predicted slots for the query that match with the characteristics of the product. 

In a typical e-commerce ranking framework, the similarity of a query with a product is composed of many features, such as the textual similarity between the query and the product description. 
We compare our proposed methods against Okapi Best Matching 25 (BM25)~\cite{robertson1995okapi, robertson1994some}, which is a text similarity measure extensively used by search engines. 
We also evaluate the retrieval performance of our methods after incorporating textual similarity of the query with the product description while calculating the similarity between a query and a product. 
To incorporate BM25 into our ranking function, we simply normalize the BM25 scores between a query and all the products between zero and one, and add the resultant score to the slot-filling score, which is used to rank the products for a query.

To the best of our knowledge, our work is the first attempt at using engagement data for distant-supervised slot-filling in e-commerce queries. This limits availability of baselines to perform a fair evaluation.
Moreover, the most closely related unsupervised method~\cite{zhai2016query} uses described grammar to annotate the queries. The grammar developed by the authors work for only two slots (product-type and brand) as developing such grammar for more diverse slots is a non-trivial task. Thus, we limit our baselines to the BM25-based approaches.

\noindent\emph{\textbf{T2: }} This task compares the predicted slots with the manually annotated queries. The candidate slot set is observed, thus, this task evaluates how well our methods can tag each word in the query with the corresponding slot, given a set of slots to choose from.

\noindent\emph{\textbf{T3: }} This task is similar to the task T2, but the candidate slot-set $c_q$ is not observed for the task T3.  Thus, in addition to predicting the tagging performance, this task also evaluates how well our methods are able to correctly choose the candidate slot-set $c_q$. When $c_q$ is not observed, it has to be sampled from the collection $I$, which can be very large (of the order $10^{12}$ for the presented experimental setup). This makes calculating Equations~\eqref{eq:c_est} and~\eqref{eq:ss_est} computationally expensive. Hence, we heuristically decrease the number of candidate slot-sets. 
For each query-word $v_{q,i}$, we find the top-$t$ most probable slots for each possible slot-key that may have generated $v_{q,i}$, i.e., top-$t$ indices corresponding to $\psi_{*,v_{q,i}}$ for each possible slot-key. The collection of the candidate slot-sets for a query then becomes equal to the all possible candidate slot-sets based on top-$t$ slots for each query word. In this paper, we use $t=1$ for simplicity.
    
Note that a query word can have only one slot-key but multiple slot values leading to annotation ambiguity. 
To avoid this ambiguity, we relax our evaluation to predict the correct slot-keys for the tasks T2 and T3. However, when the candidate slot-set is observed, predicting the correct slot-key indeed corresponds to predicting the correct slot, as there is only one possible slot-value for a slot-key. 

We generated our training, test and validation datasets from the \emph{engagement data} and \emph{annotated data} as follows:
\begin{itemize}[leftmargin=*]
    \item From the \emph{annotated data}, we used the queries not present in the engagement data to evaluate our methods on the task T3. We divided these queries in validation and test set, containing $454$ and $2{,}976$ queries, respectively. 
    \item We used the queries present in both \emph{annotated data} and \emph{engagement data} as the test dataset for evaluation on the tasks T1 and T2, leading to a total of $1{,}699$ test queries. For  T2, we generated the candidate slot-set of a query as the characteristics of the product that had the largest number of orders for the query. An additional slot \emph{miscellaneous} was added to each candidate slot-set. We use the purchase count of a product as as its relevance for a query, for  T1.
    \item We used the remaining queries from the \emph{engagement data} to estimate our models. Unlike the test data for the task T2 we prepared earlier, we do not generate $c_q$ from the most purchased item, but create a new $q, c_q$ pair for each engaged item. We added an additional slot \emph{miscellaneous} to each candidate slot-set. The size of our training set is $108{,}101$ $(q,c_q)$ pairs, with $47{,}086$ distinct queries.
\end{itemize}

\subsection{Performance Assessment metrics}\label{metrics}
For T1, we calculate the Mean Reciprocal Rank (MRR)~\cite{voorhees1999trec} and Normalized Discounted Cumulative Gain (NDCG)~\cite{wang2013theoretical} which are popular measures of the ranking quality in information retrieval.  To calculate the MRR, we assume that the relevant products for a query are the most purchased products for that query. 
To calculate the NDCG, we assume that the relevance of a product for a query is the number of times that product was purchased after issuing that query. We calculate NDCG only till rank 10 (NDCG@10), as  users usually pay attention to the top few results and break ties as proposed in~\cite{mcsherry2008computing}.

For T2 and T3, our first metric, \emph{accuracy}, calculates the overall percentage of the query words whose slot-key is correctly predicted. Our second metric calculates the percentage of the words whose slot-key is predicted correctly in a query. We report average of this percentage over all the queries and call this metric $\mqta$. Metric $\mqta$ treats all queries equally but $\ta$ is biased towards longer queries. Compared to $\ta$, $\mqta$ being a transaction-level metric gives a better picture of how our methods impact the traffic.
The metrics $\ta$ and $\mqta$ are biased towards the frequently occurring slot-keys. Therefore, we also report macro-averaged precision ($\avgprec$), recall ($\avgrec$) and F1-score ($\avgef$). 

\subsection{Parameter selection}


We used grid search on validation set to select our hyper-parameters (priors for all probability distributions, number of product categories and the parameter $\gamma$). For simplicity, we present out results for uniform Dirichlet priors, i.e., $\bm{\alpha}_i = \alpha$, $\bm{\beta}_{i,j} = \beta$, $\bm{\delta}_{i,j} = \delta$, $\bm{\zeta}_{i,j} = \zeta$, $\forall i,j$. For estimating all our approaches, we ran $1{,}000$ training iterations. We ran $100$ sampling iterations for inference of CUSDSS. We select the hyperparameters for each metric ($\ta$, $\mqta$, $\avgprec$, $\avgrec$, $\avgef$) independently, corresponding to the best performance of the metric on the validation set. Note that we perform validation only for the task T3, i.e., when $c_q$ is not observed, but we report the results for the task T2 on the same selected hyperparameters. For T1, we use the same hyperparameters corresponding to the $\mqta$ metric for T3. 

\section{Results and Discussion}\label{results}
\begin{table}[!t]
\small
\centering
  \caption{Retrieval results (Task T1).}
  \begin{threeparttable}
      \begin{tabularx}{\columnwidth}{XXRR}
        \hline
        Model & Features & MRR & NDCG@10\\\hline
         & Only BM25 & $0.036$ & $0.039$\\\hline
        USD & Only Slots & $0.071$ & $0.096$\\
            & Slots+BM25 & $0.088$ & $0.121$\\  \hline
        MSD & Only Slots & $0.072$ & $0.097$\\
            & Slots+BM25 & $0.088$ & $0.121$\\  \hline
        CUSD & Only Slots & $0.071$ & $0.096$\\
            & Slots+BM25 & $\bm{0.091}$ & $0.123$\\  \hline
        CUSDSS\textsuperscript{1} & Only Slots & $\bm{0.075}$ & $\bm{0.100}$\\
            & Slots+BM25 & $\bm{0.090}$ & $\bm{0.125}$\\  \hline
      \end{tabularx}
      \begin{tablenotes}
         \item[1] The performance of CUSDSS is statistically significant over the other models on the NDCG metric, and over USD and MSD on the MRR metric ($\pval \le 0.01$ using $\ttest$), for both the feature settings (only slots and slots+BM25).
      \end{tablenotes}
      \end{threeparttable}
  \label{tab:res_ranking}
\end{table}
\noindent\emph{\textbf{Task T1: }} Table~\ref{tab:res_ranking} shows the performance of the different approaches for the retrieval task. 
CUSDSS performs the best of all the models on both metrics, whereas CUSD performs better than USD and MSD when it is combined with BM25. This validates our hypothesis that modeling co-occurrence and subset-selection leads to better performance on slot-filling, which in turn leads to better retrieval. The performance improvement of CUSDSS over USD, using only the predicted slots in the ranking function, is $\approx5.6\%$ and $\approx4.2\%$ on the MRR and NDCG metrics, respectively. Upon incorporating the BM25 with the predicted slots in the ranking function, the performance improvement is $\approx2.3\%$ and $\approx3.3\%$ on the MRR and NDCG metrics, respectively.

On both the MRR and NDCG metrics, using the predicted slots to rank the products leads to better performance than using BM25. Additionally, using BM25 in conjunction with the predicted slots lead to even better performance. For example, the performance improvement of CUSDSS over BM25, using only the slots in the ranking function, is $\approx108\%$ and $\approx156\%$ on the MRR and NDCG metrics, respectively. The performance further improves upon incorporating BM25 with the predicted slots in the ranking function. This is particularly appreciable because the proposed approaches do not require any labeled data, instead they leverage the readily available search query logs to boost the retrieval performance. We also see that the USD and MSD performs almost equally. This is expected as all the queries \emph{nike mens running shoes}, \emph{mens nike running shoes}, \emph{running mens shoes nike} etc. are possible, favoring uniform transition probabilities. 


\begin{table}[!t]
\small
\centering
  \caption{Slot-filling results when $c_q$ is observed (Task T2).}
  \begin{threeparttable}
      \begin{tabularx}{\columnwidth}{Xrrrrr}
        \hline
        Model    & $\ta$  & $\mqta$  & $\avgprec$  & $\avgrec$ & $\avgef$\\ \hline
        USD/CUSD&$\bm{0.678}$&$\bm{0.708}$&$0.426$&$\bm{0.587}$&$0.453$\\
        MSD&$\bm{0.679}$&$\bm{0.709}$&$0.428$&$\bm{0.589}$&$0.455$\\
        CUSDSS\textsuperscript{1}&$0.657$&$0.686$&$\bm{0.571}$&$0.493$&$\bm{0.465}$\\\hline
      \end{tabularx}
      \begin{tablenotes}
         \item[1] The performance of CUSDSS is statistically significant over USD and MSD on the $\avgprec$ and $\avgef$ metrics ($\pval \le 0.01$ using $\ttest$).
      \end{tablenotes}
      \end{threeparttable}
  \label{tab:res_obs}
\end{table}

\begin{table}[!t]
\small
\centering
  \caption{Slot-filling results when $c_q$ is unobserved (Task T3).}
  \begin{threeparttable}
  \begin{tabularx}{\columnwidth}{Xrrrrr}
    \hline
    Model    & $\ta$  & $\mqta$  & $\avgprec$  & $\avgrec$ & $\avgef$\\ \hline
    USD&$0.558$&$0.583$&$0.506$&$0.645$&$0.472$\\
    MSD&$0.558$&$0.583$&$0.506$&$0.645$&$0.472$\\
    CUSD\textsuperscript{1}&$\bm{0.565}$&$\bm{0.588}$&$0.510$&$\bm{0.650}$&$0.476$\\
    CUSDSS\textsuperscript{2}&$\bm{0.566}$&$0.586$&$\bm{0.526}$&$0.622$&$\bm{0.511}$\\\hline
  \end{tabularx}
      \begin{tablenotes}
         \item[1] The performance of CUSD is statistically significant over USD and MSD on the $\ta$, $\mqta$ and $\avgrec$ metrics ($\pval \le 0.01$ using $\ttest$).
         \item[2] The performance of CUSDSS is statistically significant over USD and MSD on the $\ta$, $\mqta$, $\avgprec$ and $\avgef$ metrics and  over CUSD on the $\avgprec$ and $\avgef$ metrics ($\pval \le 0.01$ using $\ttest$)
      \end{tablenotes}
      \end{threeparttable}
  \label{tab:res_unobs}
\end{table}

\begin{figure}
\includegraphics[width=0.9\linewidth]{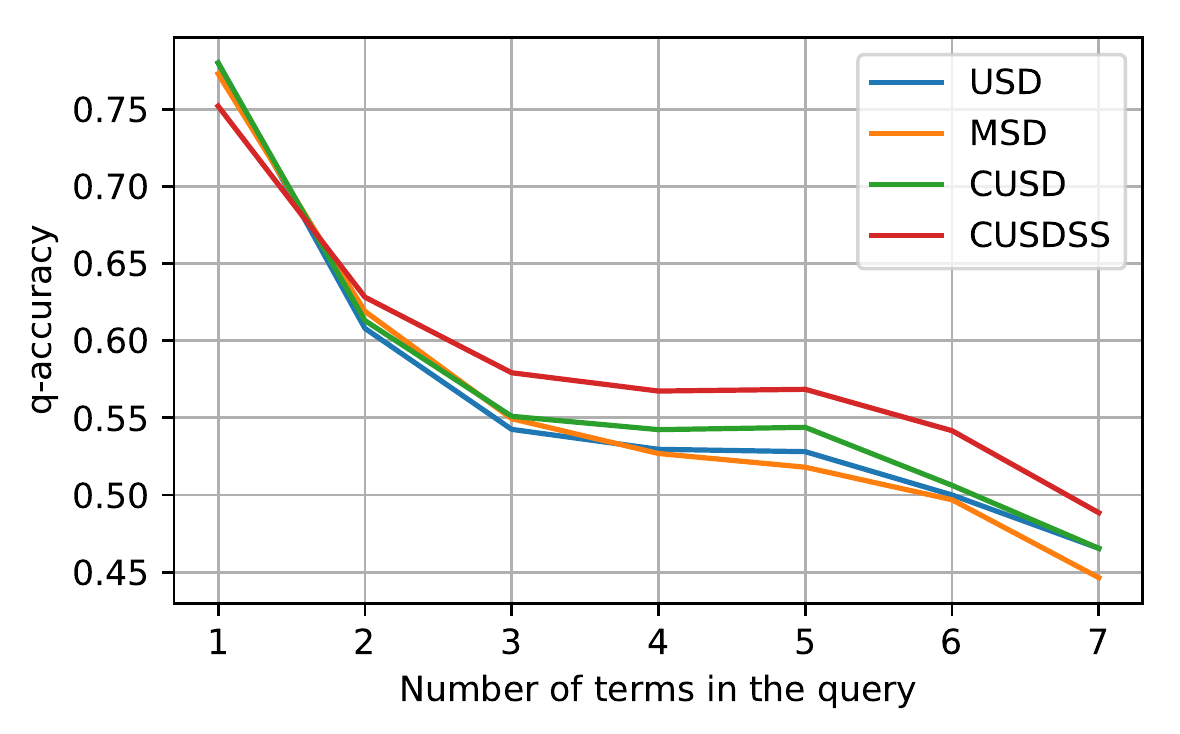}
\caption{Variation of the $q$-accuracy with query length.}
\label{fig:qac_length}
\end{figure}

\begin{table*}[!t]
\small
\centering
  \caption{Individual precision (P), recall (R) and F1-score (F1) when $c_q$ is observed.}
  {\setlength{\tabcolsep}{0.45em}\begin{tabularx}{\textwidth}{Xccccccccccccccccccccc}
    \hline
    &\multicolumn{21}{c}{Slots}\\\hline
    &\multicolumn{3}{c}{product-type}&\multicolumn{3}{c}{brand}&\multicolumn{3}{c}{gender}&\multicolumn{3}{c}{color}&\multicolumn{3}{c}{age}&\multicolumn{3}{c}{size}&\multicolumn{3}{c}{miscellaneous}\\
    Model    & P&R&F1  & P&R&F1  & P&R&F1  & P&R&F1  & P&R&F1  & P&R&F1  & P&R&F1 \\ 
    \cmidrule(lr){2-4}\cmidrule(lr){5-7}\cmidrule(lr){8-10}\cmidrule(lr){11-13}\cmidrule(lr){14-16}\cmidrule(lr){17-19}\cmidrule(lr){20-22}
    USD/CUSD&$0.93$&$0.74$&$0.82$&$0.35$&$0.89$&$0.51$&$0.63$&$0.82$&$0.69$&$0.11$&$0.64$&$0.16$&$0.27$&$0.49$&$0.36$&$0.28$&$0.27$&$0.30$&$0.42$&$0.25$&$0.33$\\
    MSD&$0.93$&$0.74$&$0.82$&$0.35$&$0.89$&$0.51$&$0.63$&$0.82$&$0.69$&$0.11$&$0.64$&$0.17$&$0.27$&$0.50$&$0.36$&$0.29$&$0.27$&$0.30$&$0.42$&$0.26$&$0.33$\\
    CUSDSS&$0.90$&$0.74$&$0.81$&$0.26$&$0.88$&$0.40$&$0.88$&$0.69$&$0.78$&$0.34$&$0.54$&$0.41$&$0.72$&$0.19$&$0.29$&$0.41$&$0.18$&$0.25$&$0.49$&$0.24$&$0.32$\\\hline
  \end{tabularx}}
  \label{tab:slots_obs}
\end{table*}
\begin{table*}[!t]
\small
\centering
  \caption{Individual precision (P), recall (R) and F1-score (F1) when $c_q$ is unobserved.}
  {\setlength{\tabcolsep}{0.45em}\begin{tabularx}{\textwidth}{Xccccccccccccccccccccc}
    \hline
    &\multicolumn{21}{c}{Slots}\\\hline
    &\multicolumn{3}{c}{product-type}&\multicolumn{3}{c}{brand}&\multicolumn{3}{c}{gender}&\multicolumn{3}{c}{color}&\multicolumn{3}{c}{age}&\multicolumn{3}{c}{size}&\multicolumn{3}{c}{miscellaneous}\\
    Model    & P&R&F1  & P&R&F1  & P&R&F1  & P&R&F1  & P&R&F1  & P&R&F1  & P&R&F1 \\ 
    \cmidrule(lr){2-4}\cmidrule(lr){5-7}\cmidrule(lr){8-10}\cmidrule(lr){11-13}\cmidrule(lr){14-16}\cmidrule(lr){17-19}\cmidrule(lr){20-22}
    USD&$0.91$&$0.59$&$0.71$&$0.20$&$0.85$&$0.35$&$0.81$&$0.93$&$0.76$&$0.39$&$0.88$&$0.42$&$0.17$&$0.62$&$0.42$&$0.23$&$0.50$&$0.40$&$0.83$&$0.14$&$0.24$\\
    MSD&$0.91$&$0.59$&$0.71$&$0.20$&$0.85$&$0.35$&$0.81$&$0.93$&$0.77$&$0.39$&$0.88$&$0.41$&$0.17$&$0.62$&$0.42$&$0.23$&$0.50$&$0.40$&$0.83$&$0.14$&$0.23$\\
    CUSD&$0.91$&$0.60$&$0.72$&$0.20$&$0.85$&$0.36$&$0.81$&$0.94$&$0.76$&$0.40$&$0.88$&$0.42$&$0.18$&$0.63$&$0.43$&$0.24$&$0.51$&$0.41$&$0.84$&$0.14$&$0.24$\\
    CUSDSS&$0.90$&$0.61$&$0.72$&$0.19$&$0.84$&$0.32$&$0.83$&$0.90$&$0.86$&$0.37$&$0.87$&$0.52$&$0.46$&$0.53$&$0.48$&$0.34$&$0.41$&$0.38$&$0.61$&$0.20$&$0.30$\\\hline
  \end{tabularx}}
  \label{tab:slots_unobs}
\end{table*}

\begin{table*}[!t]
\small
\centering
  \caption{Examples of product categories estimated by CUSD and CUSDSS}
  {\setlength{\tabcolsep}{0.45em}\begin{tabularx}{\textwidth}{Xp{7.5cm}p{7.5cm}}
  \hline
    Product category    & Most probable slots estimated by CUSD (ranked by probability)  & Most probable slots estimated by CUSDSS (ranked by probability) \\
    \hline
    Houseware    & age: adult, gender: unisex, brand: mainstays, size: 1, color: white, color: black, brand: better homes \& gardens, color: brown, pt: storage chests \& boxes, brand: sterilite, color: silver, color: gray& brand: mainstays, pt: storage chests \& boxes, pt: vacuum cleaners, brand: sterilite, brand: better homes \& gardens, brand: the pioneer woman, pt: food storage jars \& containers, brand: holiday time\\\hline
    Toys    & pt: dolls, color: multicolor, age: child, gender: unisex, brand: barbie, pt: doll playsets, gender: girls, age: adult, brand: shopkins, brand: disney princess, brand: my life as, age: 3 years \& up, brand: disney&pt: dolls, pt: action figures, pt: play vehicles, brand: barbie, pt: doll playsets, pt: stuffed animals \& plush toys, brand: disney, pt: interlocking block building sets, brand: lego, brand: paw patrol\\\hline
    Dental hygiene   & gender: unisex, age: adult, color: multicolor, pt: toothpastes, brand: crest, brand: colgate, brand: oral-b, size: 1, color: white, age: child, size: 4, pt: mouthwash, pt: powered toothbrushes, age: teen, size: 6&pt: toothpastes, brand: crest, brand: colgate, brand: oral-b, pt: baby foods, brand: gerber, pt: mouthwash, pt: meal replacement drinks, pt: manual toothbrushes, pt: powered toothbrushes, brand: ensure\\\hline
  \end{tabularx}}
  \label{tab:pt_cats}
\end{table*}
\noindent\emph{\textbf{Task T2 and T3: }} Tables~\ref{tab:res_obs} and~\ref{tab:res_unobs} show the performance for the slot classification when the candidate slot-set is observed and unobserved, respectively.
We report average of $100$ runs with  random initializations.
For all the methods, we see that $\mqta$ is more than $\ta$, which shows that all the methods perform better on shorter queries. Figure ~\ref{fig:qac_length} shows how the $\mqta$ varies with the number of words in the queries. For all the methods, $\mqta$ decreases with the length of query. Similar to the ranking task T1, we see that USD and MSD performs equally good for both the cases. CUSD performs better than USD and MSD on all the metrics, which shows that selecting $c_q$ with co-occurring slots leads to better performance. Moreover, co-occurrence information is exploited when the query has multiple query words. As shown in Figure \ref{fig:qac_length}, CUSD performs considerably better than USD and MSD on the $\mqta$ metric as the number of words in the queries increases.  

Interestingly, CUSDSS performs better than the USD and MSD on the $\mqta$ and $\ta$ metrics when the candidate slot-set is observed, but does not perform as good when the candidate slot-set is unobserved. In addition, CUSDSS performs best on the $\avgprec$ and $\avgef$ metrics, but not so good on the $\avgrec$ metrics. To investigate this peculiar behavior of CUSDSS, we compared the individual precision, recall and F1 scores for the slot-keys of CUSDSS with the other approaches. Tables~\ref{tab:slots_obs} and~\ref{tab:slots_unobs} shows these statistics for each of the slot when the candidate slot-set is observed and unobserved, respectively. Subset selection leads to CUSDSS labeling more query words with frequent slots like \emph{product-type} and \emph{brand}. This not only decreases the recall of the other less frequent slots (\emph{gender, color, age, size}) but also decreases the precision of the frequent slots. However, this also tends to make CUSDSS predict only those less frequent slots for which it has high confidence, leading to an increase on the $\avgprec$ metric. Consequently, CUSDSS also performs the best on the $\avgef$ metric ($3\%$ over the USD when $c_q$ is observed, and $8\%$ improvement when $c_q$ is not observed). When the candidate slot-set is unobserved, there are more possible ways to make an error in slot-prediction, than the case when the candidate slot-set is observed. For the unobserved case, as the prediction is biased towards the many slots, CUSDSS performs better on the micro-averaged metrics, i.e., $\ta$ and $\mqta$. However, for observed candidate slot, CUSDSS does not benefit as the scope of error is small.

Additionally, even though the overall performance of CUSDSS on the $\mqta$ metric is not at par with the CUSD, the performance is considerably better for the queries with multiple words, and the difference in performance increases with the number of words in the query, as depicted in Figure \ref{fig:qac_length}. When there is only one word in the query, CUSDSS favors frequent slots, leading to degraded performance as there is no co-occurrence information to be leveraged.

Te study qualitatively how subset selection helps CUSDSS make better predictions than CUSD. Table \ref{tab:pt_cats} shows the most probable slots for some product categories estimated by CUSDSS and CUSD. We manually label the estimated product categories with a relevant semantic name. We observe that the product categories estimated by CUSD contain more slots corresponding to the slot-keys \emph{age, gender, size} and \emph{color}, while the product categories estimated by CUSDSS contain more slots corresponding to the slot-keys \emph{product-type} and \emph{brand}. For example, the product category \emph{houseware} for CUSD has \emph{age: adult} as the most probable slot, while CUSDSS has \emph{brand: mainstays} as the most probable slot. 
This follows from the fact that CUSD models the complete candidate slot-sets $c_q$ as belonging to the same product category, which leads to the frequent occurring slots in $c_q$ having high probabilities in the product categories. In contrast, CUSDSS models the actual product-intent $a_q$ as belonging to the same product category, which leads to the actual intended slots having high probabilities in the product categories.

\section{Conclusion}\label{conclusion}

In this paper, we used the historical query reformulation logs of e-commerce retailers to develop distant-supervised approaches to perform slot-filling for the e-commerce queries. Our approaches solely depend on the user engagement data and require no manual labeling effort. 
In terms of retrieval, our approaches achieve better ranking performance (up to $156\%$) over Okapi BM25. In addition, when used along with Okapi-BM25, our approaches improve the ranking performance by $250\%$. 
Our work makes a step towards leveraging distant-supervision for slot-filling, and envision that the proposed method will serve as a motivation for other applications that rely on the labeled training data, which is expensive and time-consuming. To our best knowledge, this paper is the first attempt in only relying upon the engagement data to perform slot-filling for the e-commerce queries. Code accompanying with this paper will be available at Github.




\bibliographystyle{ACM-Reference-Format}
\bibliography{sample-bibliography}

\end{document}